\documentclass[12pt]{article}

\usepackage{amssymb}

\begin{document}

\title{A stationary vacuum solution dual to the Kerr solution}

\author{Z. Ya. Turakulov\thanks{\n 
Permanent address: \, Institute  of  Nuclear Physics, \, Ulugbek, \, Tashkent 
\, 702132, \, Uzbekistan. \hspace*{0.5cm} E-mail: zafar@suninp.tashkent.su} \,
 and 
N. Dadhich \thanks{E-mail: nkd@iucaa.ernet.in}\\
{\sl Inter-University Centre for Astronomy \&
Astrophysics,} \\
{\sl Post Bag 4, Ganeshkhind,
Pune 411 007, India .}}

\date{}
\newcommand{\gcc}{$g~cm^{-3}\ $}
\newcommand{\sfun}[2]{$#1(#2)\ $}
\newcommand{\rhonot}{$\rho_{\circ}\ $}
\newcommand{\msun}{$M_{\odot}\ $}
\newcommand{\greq}{$\stackrel{>}{ _{\sim}}$}
\newcommand{\lteq}{$\stackrel{<}{ _{\sim}}$}
\def\th{\theta}
\def\n{\noindent}

\maketitle
\begin{abstract}
We present a stationary axially symmetric two parameter vacuum solution 
which could be considered as ``dual'' to the Kerr solution. It is obtained 
by removing the mass parameter from the function of the radial coordinate 
and introducing a dimensionless parameter in the function of the angle 
coordinate in the metric functions. It turns out that it is in fact the 
massless limit of the Kerr - NUT solution.
\end{abstract}


 The axially symmetric stationary spacetime hosts the most interesting
two parameter Kerr family of vacuum black hole solution of the
Einstein equation. It is well known that the family is unique under the
assumptions of asymptotic flatness and of existence of regular smooth
horizon. The Kerr solution has turned out to be the most interesting solution
 of the Einstein equation for the astrophysical applications, particularly in 
the context of high energy sources like quasars, pulsars, gamma ray bursters 
and active galactic nuclei. 

 The two parameters in the solution represent mass and specific angular 
momentum of 
the hole. When rotation is switched off, the solution reduces to the static 
Schwarzschild black hole. In the limit of vanishing mass it reduces to flat 
space. That is, the  rotation by itself cannot be a source of gravity. This 
is however quite understandable in the Newtonian framework. The question is, 
should that always be the case in general relativity as well? In this paper, 
we wish to address this question and would like to present a solution in 
which mass parameter is replaced by a dimensionless parameter 
appropriately. The solution so obtained would in a particular sense be 
``dual'' to the Kerr solution. 

 The form of the axially symmetric stationary metric we choose is motivated 
by the physical considerations of separability of the Hamilton - Jacobi 
equation for particle motion and the Klein - Gordon equation. That is, we seek
the solution of the Einstein vacuum equation in the metric form in which 
these equations characterizing the physical features of the solution are 
solvable. For this, we follow the method employed by one of us (ZT) [1]. This 
would determine the form and character of the certain metric functions, a 
priori, and the metric is written in the following form [1],
\begin{eqnarray}
ds^2 &=& \nonumber \Lambda ( dt + a\frac{(F - U^2)V^2 - (G - V^2)U^2}{U^2 
- a^2V^2} d\varphi)^2 - \Lambda^{-1}[(U^2 - a^2V^2)(\frac{dr^2}{U^2} + 
\frac{dS^2}{V^2}) + U^2V^2 d\varphi^2] \\
\end{eqnarray}
where
\begin{equation}
\Lambda = \frac{U^2 - a^2V^2}{F - a^2G}, 
\end{equation}
\begin{equation}  
F= r^2 + a^2, \, ~G = 1 - S^2. 
\end{equation}

 Further we have $U = U(r), V = V(S)$ where $S$ is an angle coordinate and 
$a$ is a constant having the dimension of length. 

Now the Kerr solution is specified by
\begin{equation}
F - U^2 = 2Mr, ~G - V^2 = 0
\end{equation}
with $M$ and $a$ having the usual meaning of mass and specific angular 
momentum of the rotating black hole. On the other hand, the specification
\begin{equation}
F - U^2 = 0, ~G - V^2 = -2NS
\end{equation}
also gives a vacuum solution and 
where the new parameter $N$ is dimensionless. This characterization is a kind 
of ``dual''. In the former the radial part $F - U^2$, which is dimensionful, 
is specified through the dimensionful mass parameter and the angle part 
$G - V^2$ is vacuous while in the latter the radial part is vacuous and the 
angle part, which is dimensionless, is specified through a dimensionless 
parameter $N$. In this sense the above two characterizing relations are 
``dual'' to each-other, and so should be the spacetimes they specify.

 Clearly, it is this form of the metric which has suggested us the duality 
relation leading to the dual solution. It is an 
example of pure gravomagnetic spacetime because the primary source of the 
field is introduced through the angle function in eqn. (4). Note that what is 
the radial coordinate to gravoelectric and so is the angle coordinate to 
gravomagnetic spacetime. The axially symmetric vacuum spacetime metric 
in the proper (separability of H-J and K-G equations) form (1) is 
characterized by the two functions $U(r)$ and 
$V(S)$. The mass parameter enters through the 
radial function $U$ while the dimensionless parameter through the angular 
function $V$.

 Further, the dual character of the parameters $M$ and $N$ will  become 
clearly visible in the expressions for the  curvature components. For instance,
 let us write one particular component, say $R^{rS}{}_{rS}$ for the metric (1)
 explicitly,
\begin{equation}
R^{rS}{}_{rS} = \frac{Mr(r^2 - 3a^2S^2)}{(r^2 + a^2S^2)^3}
\end{equation}
for the Kerr solution specified by eqn. (3). This component would go over to 
the one for the specification of the dual solution (4) by the transformation,
$M\rightarrow -aN, ~r\leftrightarrow aS$. This is a true dual transformation 
 which takes the Kerr solution to the dual solution;i.e. 
Riemann(Kerr) $\rightarrow$ Riemann(Dual). Under this transformation, the 
functions of $r$ and $S$ interchange their roles, $(F - U^2)\leftrightarrow 
a^2(G - V^2)$. Then eqns (3) and (4) are clearly the dual of each-other and 
so are the spacetimes described by them.

 For the dual solution, the metric would take the explicit form,
\begin{eqnarray}
ds^2 &=& \nonumber \frac{r^2 + a^2\cos^2\th - a^2N^2\sin^2\th}{r^2 + a^2S^2} 
(dt + \frac{2aNS(r^2 + a^2)}{r^2 + a^2\cos^2\th - a^2N^2 \sin^2\th} 
d\varphi)^2  \\ 
&-& \frac{(1 + N^2)(r^2 + a^2)(r^2 + a^2S^2)\sin^2\th}{r^2 + a^2\cos^2\th 
- a^2N^2\sin^2\th} d\varphi^2 
 - (r^2 + a^2S^2)(\frac{dr^2}{r^2 + a^2} + d\th^2)
\end{eqnarray}
where we have written $S = N + (1 + N^2)^{1/2}\cos\th$.

 It reduces to flat space when $N = 0$ or $a = 0$. The spacetime is 
asymptotically non flat. It turns out that it is in fact the Kerr - NUT 
solution [2] with $M = 0$. That is, it is the massless limit of the Kerr - NUT 
solution. By writing $l = aN, a^2 = b^2 - l^2$ (and then replacing $b$ by $a$) 
and redefining the coordinates 
$t$ and $\varphi$ appropriately, the above metric can be transformed to the 
Kerr - NUT form,
\begin{eqnarray}
ds^2 &=& \nonumber \frac{A}{B}[dt - (a \sin^2\th - 2 l \cos\th) d\varphi]^2 
 - \frac {B}{A}dr^2  \\ 
&-& \frac{\sin^2\th}{B}[(r^2 + a^2 + l^2)d\varphi - a dt]^2 - B d\th^2
\end{eqnarray}
where
\begin{equation}
A = r^2 + a^2 - l^2,  ~B = r^2 + (l + a \cos\th)^2 .
\end{equation}

 This is the Kerr - NUT solution [2] with $M = 0$ and $l$ being the NUT 
parameter. The massless limit of the 
Kerr - NUT solution and the Kerr solution are thus dual of each-other. The NUT 
solution [3-7] is the prime example of the gravomagnetic field, in particular 
 it can be interpreted as the field of a grvomagnetic monopole charge [4]. In 
 our solution, though $M=0$, yet the Kerr and NUT parameters do contribute 
an effective gravitational mass $(l^2 - a^2)/r$. This follows from writing 
$A/r^2 = 1 + 2\phi$, where $\phi$ is the Newtonian potential. The effective 
mass would be positive or negative depending upon $l^2$ $\gtrless$ $ a^2$ and 
consequently the field would be attractive or repulsive. That is, the NUT 
parameter produces attractive field while the Kerr parameter, as is 
well - known, produces repulsive. 

 The maim aim of this exercise was to expose this interesting duality 
relation between the two exact vacuum solutions. This duality is however 
different from the elctrogravity duality considered in Refs. [8,9]. The latter 
referred to duality between active and passive electric parts of the 
Riemann curvature, which was also the symmetry of the vacuum equation. Here we 
are instead  referring to duality in the prescription of the radial and 
angular functions in the metric.

 Acknowledgment: We wish to thank our young colleague Parampreet Singh for 
helping us in verifying the solution. It is a pleasure to thank Dharamveer 
Ahluwalia for useful suggestions. ZT wishes to thank ICTP, Trieste for 
the travel grant under its BIPTUN Programme which facilitated this 
collaboration and also thanks IUCAA for warm hospitality. This result was 
first presented by one of us (ND) in the Symposium on 
{\it Conceptual Issues in Relativity, Astrophysics and Cosmology }, 
March, 28-30, North Bengal University, Siliguri.

\end{document}